# Influence of secondary gas on control characteristics of ICP-RF plasma torches

P J Bhuyan*[1,2] & K S Goswami[2]

[1]Department of Physics, Tezpur University
Napaam, Tezpur, Assam. PIN -784028 (India)
[2]Centre of Plasma Physics- Institute for plasma research,
Nazirakhat, Sonapur, Assam, Pin - 782402(India)

*E-mail: pranjal.bhuyan@rediffmail.com, pjb@tezu.ernet.in

**Abstract**
Thermal plasma characteristics inside an atmospheric pressure Radio-Frequency (RF) Argon - plasma torch have been studied by numerically solving axis-symmetric 2-D MHD equations, energy transport equations and species conservation equation coupled with 2D Maxwell's equations in the vector potential form. A renormalization group (RNG) $k$-$\varepsilon$ model was employed to study the turbulence within the torch. Helium is mixed with Argon for injection as the sheath gas, and its effect on the control characteristics of the torch has been studied. A control volume approach and semi-implicit pressure linked equations revised (SIMPLER) algorithm was used to solve the above set of equations to obtain the flow, temperature, turbulence and EM source fields within the torch. Finally, a comparison of the results obtained in our present calculation has been made with other works of similar nature. The torch geometry, flow rates and power dissipation values used in the present calculation are similar to the standard ones available in literature.

## 1. Introduction

The electrodes in DC plasma [1, 2] are infact a limitation for their application to high purity material processing because of the contamination of plasma caused by eroded electrode material. In this context, the RF plasma has the advantage of large volume, low velocity, cleanliness and continuity since it is produced by an induction RF electric filed without the use of electrodes. RF induction plasma torches are now widely used in different fields, such as processing of high-purity materials, synthesis of ultrafine powders of metals, alloys and ceramics etc [1, 3-11]. An in-depth understanding of the basic plasma processes involved in the operation of RF plasma torches is essential for improving our ability to control them & ensure successful operation of these devices.

Mathematical modeling of the thermal plasma processes gives us an effective and economical tool for studying the torch control parameters, such as flow, temperature and concentration fields inside the induction plasma over a wide range of operating conditions and torch geometry. Moreover, it can be used to probe the areas within the plasma torch where the experimental diagnostics has not been feasible yet. This justifies the amount of research work that has been put into the field of atmospheric pressure [1, 3-11] as well as low pressure [12-13] RF plasma simulation and modeling over the last two decades.

In this paper, we perform a numerical study on the control parameters inside a RF Argon plasma torch operating at atmospheric pressure. We study an inductively coupled RF plasma torch (ICP-RF) with typical geometry [6] and flow conditions with Argon as the working gas. Different concentrations of Helium is mixed with Argon and introduced to the torch as the sheath gas. Diffusion of this gas into the central regions of the torch under different swirling schemes and its effect on the conditions within the torch will be investigated. Turbulence will be studied with a modified k-$\varepsilon$ model, namely the RNG $k$-$\varepsilon$ model which was earlier applied in the case of RF plasma torch under transient flow conditions [8].

## 2. Theoretical formulation

### 2.1. Torch Geometry and basic assumptions

Figure 1 shows the schematic diagram of the RF induction plasma torch and the coordinate system. A summary of torch geometry and operating conditions is tabulated in table 1. The following assumptions are made for the RF induction thermal plasma in order to introduce the governing equations [4]:

i. The plasma is optically thin continuous fluid in LTE.
ii. The flow is steady and turbulent with swirling. The velocity, temperature and concentration fields are 2-D and axis symmetric.
iii. The induction electromagnetic field is 2D and axis symmetric.
iv. The displacement current, gravity and viscous dissipation, density and temperature fluctuations are not considered.
v. The effects of anisotropy on the transport properties and on the turbulent structure are not considered.

## 2.2  Governing equation

Under the given assumptions, the governing equations read in the cylindrical coordinates $(z, r, \theta)$.

**The continuity equation:**

$$\frac{\partial}{\partial z}(\rho u) + \frac{1}{r}\frac{\partial}{\partial r}(r\rho v) = 0 \tag{1}$$

**Momentum equations:**

(i)  z- component

$$\frac{\partial}{\partial z}(\rho u u) + \frac{1}{r}\frac{\partial}{\partial r}(r\rho v u) = -\frac{\partial p}{\partial z} + F_z + 2\frac{\partial}{\partial z}\left(\mu_{eff}\frac{\partial u}{\partial z}\right) + \frac{1}{r}\frac{\partial}{\partial r}\left[r\mu_{eff}\left(\frac{\partial u}{\partial r} + \frac{\partial v}{\partial z}\right)\right] \\ - \frac{2}{3}\frac{\partial}{\partial z}\left[\mu_{eff}\left(\frac{\partial u}{\partial z} + \frac{1}{r}\frac{\partial(rv)}{\partial r}\right)\right] \tag{2}$$

(ii)  r-component

$$\frac{\partial}{\partial z}(\rho u v) + \frac{1}{r}\frac{\partial}{\partial r}(r\rho v v) = -\frac{\partial p}{\partial r} + F_r + \frac{2}{r}\frac{\partial}{\partial r}\left(r\mu_{eff}\frac{\partial v}{\partial r}\right) + \frac{\partial}{\partial z}\left[\mu_{eff}\left(\frac{\partial u}{\partial r} + \frac{\partial v}{\partial z}\right)\right] \\ - \mu_{eff}\frac{2v}{r^2} - \frac{2}{3}\frac{\partial}{\partial z}\left[\mu_{eff}\left(\frac{\partial u}{\partial z} + \frac{1}{r}\frac{\partial(rv)}{\partial r}\right)\right] + \rho\frac{w^2}{r} \tag{3}$$

(iii)  $\theta$- component

$$\frac{\partial}{\partial z}(\rho u w) + \frac{1}{r}\frac{\partial}{\partial r}(r\rho v w) = \frac{\partial}{\partial z}\left(\mu_{eff}\frac{\partial w}{\partial z}\right) + \frac{1}{r}\frac{\partial}{\partial r}\left[r\mu_{eff}\frac{\partial w}{\partial r}\right] - \rho\frac{vw}{r} - \frac{w}{r^2}\frac{\partial}{\partial r}(r\mu_{eff}) \tag{4}$$

in the above equations, $u$, $v$ and $w$ are the time-averaged axial, radial and tangential velocity components respectively, $p$ is the pressure, $\rho$ is the density and $\mu_{eff}$ is the effective viscosity, which is expressed as the sum of laminar ($\mu_l$) and turbulent ($\mu_t$) parts as

$$\mu_{eff} = \mu_l + \mu_t \tag{5}$$

with

$$\mu_t = \frac{\rho C_\mu k^2}{\varepsilon} \tag{6}$$

**Energy equation**

The conservation of energy is expressed in terms of enthalpy $h$ as:

$$\frac{\partial}{\partial z}(\rho u h) + \frac{1}{r}\frac{\partial}{\partial r}(r\rho v h) = \frac{\partial}{\partial z}\left[\left(\frac{\kappa}{C_p} + \frac{\mu_t}{\sigma_t}\right)\frac{\partial h}{\partial z}\right] + \frac{1}{r}\frac{\partial}{\partial r}\left[r\left(\frac{\kappa}{C_p} + \frac{\mu_t}{\sigma_t}\right)\frac{\partial h}{\partial r}\right] + Q_j - Q_r \tag{7}$$

Where $\kappa$ is the thermal conductivity, $C_p$ is the specific heat at constant pressure and $\sigma_t$ is the Prandtl number, $Q_j$ is the joules heating due to the RF field, and $Q_r$ is the volumetric radiation loss.

**The turbulence equations**

In the present work, a two-dimensional, time-independent renormalization group (RNG) $k$-$\varepsilon$ model for RF plasma [8] has been used. The time-independent transport equations representing this model are:

$$\frac{\partial}{\partial z}(\rho u k) + \frac{1}{r}\frac{\partial}{\partial r}(r\rho v k) = \frac{\partial}{\partial z}\left[\alpha_k \mu_{eff}\frac{\partial k}{\partial z}\right] + \frac{1}{r}\frac{\partial}{\partial r}\left[r\alpha_k \mu_{eff}\frac{\partial k}{\partial r}\right] + G - \rho\varepsilon \qquad (8)$$

$$\frac{\partial}{\partial z}(\rho u \varepsilon) + \frac{1}{r}\frac{\partial}{\partial r}(r\rho v \varepsilon) = \frac{\partial}{\partial z}\left[\alpha_\varepsilon \mu_{eff}\frac{\partial \varepsilon}{\partial z}\right] + \frac{1}{r}\frac{\partial}{\partial r}\left[r\alpha_\varepsilon \mu_{eff}\frac{\partial \varepsilon}{\partial r}\right] + \frac{\varepsilon}{k}(C_1 G - \rho C_2 \varepsilon - R) \qquad (9)$$

Here, $G$ represents the production rate of turbulent kinetic energy, given as:

$$G = \mu_t \left\{ 2\left[\left(\frac{\partial u}{\partial z}\right)^2 + \left(\frac{\partial v}{\partial r}\right)^2 + \left(\frac{v}{r}\right)^2\right] + \left(\frac{\partial w}{\partial r} - \frac{w}{r}\right)^2 + \left(\frac{\partial w}{\partial z}\right)^2 + \left(\frac{\partial v}{\partial z} + \frac{\partial u}{\partial r}\right)^2 \right\} \qquad (10)$$

The RNG $k$-$\varepsilon$ model is thus similar to standard $k$-$\varepsilon$ [6,7] model except for the term R, which has the form:

$$R = \frac{C_\mu \rho \eta^3 \left(1 - \eta/\eta_o\right)\varepsilon}{1 + \beta\eta^3} \qquad (11)$$

Where $\eta_o = 4.38$, $\beta = 0.012$ and $\eta \equiv Sk/\varepsilon$, $S$ being the modulus of mean rate of strain, which is related to G as:

$$S = \sqrt{G/\mu_t} \qquad (12)$$

The constants of the turbulence model, as proposed by Launder and Spalding are:
$C_\mu = 0.09 \qquad C_1 = 1.42 \qquad C_2 = 1.68 \qquad \alpha_k = 1.39 \qquad \alpha_\varepsilon = 1.39 \qquad \sigma_t = 0.7$

**Species conservation equation**
In cases where Helium is added with Argon and injected as sheath gas, the concentration of different species is obtained from the species conservation equation:

$$\frac{\partial}{\partial z}(\rho u C_w) + \frac{1}{r}\frac{\partial}{\partial r}(r\rho v C_w) = \frac{\partial}{\partial z}\left[\Gamma_C \frac{\partial C_w}{\partial z}\right] + \frac{1}{r}\frac{\partial}{\partial r}\left[r\Gamma_C \frac{\partial C_w}{\partial r}\right] \qquad (13)$$

With $\Gamma_C = \frac{\mu_l}{S_{cl}} + \frac{\mu_t}{S_{ct}}$. For neutral species, neglecting the presence of ions, $S_{ct} = S_{cl} = 1$.

The thermodynamic and transport properties of Ar-He mixed plasma are obtained from the mixture rule [1]. If $y_1$ and $y_2$ are the molar fractions and $M_1$ and $M_2$ are the molar masses of the two non-polar gases, then

$$\mu = \frac{y_1 \mu_1}{y_1 + y_2 \phi_{12}} + \frac{y_2 \mu_2}{y_2 + y_1 \phi_{21}} \qquad (14)$$

With

$$\phi_{12} = \frac{\left[1 + (\mu_1/\mu_2)^{1/2}(M_2/M_1)^{1/4}\right]^2}{\left[8(1 + M_2/M_1)\right]^{1/2}} \qquad (15)$$

$$\phi_{21} = \frac{\mu_2}{\mu_1}\frac{M_1}{M_2}\phi_{12} \qquad (16)$$

$$y_i = \frac{\dfrac{C_{wi}}{M_i}}{\dfrac{C_{w1}}{m_1} + \dfrac{C_{w2}}{M_2}} \qquad (17)$$

$i = 1, 2$

The thermal and electrical conductivities are also evaluated from equations similar to equation (14). Finally the specific heat and is deduced from linear interpolation [1]:

$$C_p = y_1 C_{p1} + y_2 C_{p2} \qquad (18)$$

**The vector potential equations**

The electromagnetic field effects produced by the RF field as well as the induced field within the torch is given by the formulation developed by Mostaghimi [3], that gives the real and imaginary components of vector potential as:

$$\frac{1}{r}\frac{\partial}{\partial r}\left(r\frac{\partial A_R}{\partial r}\right) + \frac{\partial^2 A_R}{\partial z^2} - \frac{A_R}{r^2} + \mu_0 \sigma \omega A_I = 0 \qquad (19)$$

$$\frac{1}{r}\frac{\partial}{\partial r}\left(r\frac{\partial A_I}{\partial r}\right) + \frac{\partial^2 A_I}{\partial z^2} - \frac{A_I}{r^2} - \mu_0 \sigma \omega A_R = 0 \qquad (20)$$

From these equations, we calculate the electric and magnetic fields associated with the flow as:

$$E_\theta = -i\omega A_\theta \qquad (21)$$

$$\mu_0 H_z = \frac{1}{r}\frac{\partial}{\partial r}(rA_\theta) \qquad (22)$$

$$\mu_0 H_r = -\frac{\partial}{\partial z}(A_\theta) \qquad (23)$$

Finally, the forces and Joules heating terms appearing in the momentum and energy conservation equations are given in terms of the quantities calculated in equations (21)-(23) as:

$$F_z = -\frac{1}{2}\mu_0 \sigma \operatorname{Re}(E_\theta \overline{H}_r) \qquad (24)$$

$$F_r = \frac{1}{2}\mu_0 \sigma \operatorname{Re}(E_\theta \overline{H}_z) \qquad (25)$$

$$Q_j = \frac{1}{2}\sigma E_\theta . \overline{E}_\theta \qquad (26)$$

**3. Boundary conditions**

**At the Entrance (z = 0)**

$T_o = 350$ K

$$u = \begin{cases} \dfrac{Q_1}{\pi r_1^2} = u_1 & r < r_1 \\ 0 & r_1 \leq r \leq r_2 \\ \dfrac{Q_2}{\pi(r_3^2 - r_2^2)} & r_2 \leq r \leq r_3 \\ \dfrac{Q_3}{\pi(R_o^2 - r_3^2)} & r_3 \leq r \leq R_o \end{cases}$$

$v = 0$

$$w = \begin{cases} 0 & r < r_2 \\ w_2 & r_2 \leq r \leq r_3 \\ w_3 & r_3 \leq r \leq R_o \end{cases}$$

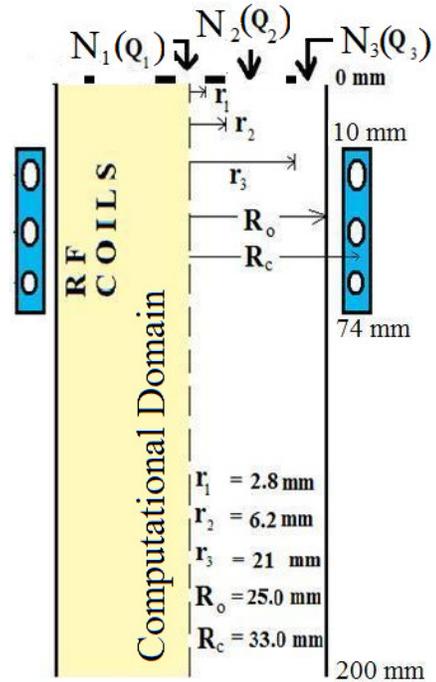

$r_1 = 2.8$ mm
$r_2 = 6.2$ mm
$r_3 = 21$ mm
$R_o = 25.0$ mm
$R_c = 33.0$ mm

*Figure 1- Geometry of the Torch*

$$Cw = \begin{cases} 0 & r < r_2 \\ 0 & r_2 \leq r \leq r_3 \\ C_w & r_3 \leq r \leq R_o \end{cases}$$

$$\frac{\partial^2 A_R}{\partial z^2} = \frac{\partial^2 A_I}{\partial z^2} = 0$$

$k = 0.005\,(u^2 + w^2)$

$\varepsilon = 0.1\,k^2$

**At the exit** (z = 200 mm)

$$\frac{\partial(\rho u)}{\partial z} = \frac{\partial v}{\partial z} = \frac{\partial w}{\partial z} = 0$$

$$\frac{\partial h}{\partial z} = \frac{\partial k}{\partial z} = \frac{\partial \varepsilon}{\partial z} = \frac{\partial^2 A_R}{\partial z^2} = \frac{\partial A_I}{\partial z^2} = 0$$

**At the centerline** (r = 0)

$$v = w = \frac{\partial u}{\partial r} = 0$$

$$\frac{\partial h}{\partial r} = \frac{\partial k}{\partial r} = \frac{\partial \varepsilon}{\partial r} = A_R = A_I = 0$$

**At the wall** ($r = R_o$)

$u = v = w = 0$

$k = \varepsilon = 0$

Table I

| Property | Value |
| --- | --- |
| $Q_1$ (l min$^{-1}$) | 6.3 |
| $Q_2$ (l min$^{-1}$) | 30 |
| $Q_3$ (l min$^{-1}$) | 150 |
| $f$ (M Hz) | 3 |
| $\kappa_c$ (W m$^{-1}$ K$^{-1}$) | 1.047 |
| $\delta_w$ (mm) | 2 |
| $P_o$ (kW) | 8 |

*Table 1- Operating parameters of the torch*

For the grid points near the wall, the value of turbulent kinetic energy and its dissipation rates, i.e. $k_w$ and $\varepsilon_w$ are deduced by the wall functions from the following expression:

$$k_w = \frac{u_\tau^2}{\sqrt{C_\mu}}$$

and

$$\varepsilon_w = \frac{u_\tau^3}{K_a y_p}$$

Where $K_a$ is the von Karman constant ($K_a = 0.41$) and $y_p$ is the first mesh point away from the wall in the computational domain. The frictional velocity $u_\tau$ is calculated from:

$$\frac{u_w}{u_\tau} = \frac{1}{K_0} \ln\left(\frac{u_\tau y_p}{\mu}\right) + B$$

Where $u_w$ is the axial velocity at $y_p$ and B is a constant (B = 5.0)

Finally, the temperature near the wall is determined from,

$$k_{eff} \frac{\partial T}{\partial r} = \frac{\kappa_c}{\delta_w}(T - T_w)$$

Where $\delta_w$ is the tube thickness $\kappa_c$ is the thermal conductivity of the Plexiglass, and $T_w = 350$ K is the temperature of the outer surface of the tube.

Finally, the boundary condition on Vector potentials is given by:

$$A_R = \frac{\mu_0 I}{2\pi}\left(\frac{r_c}{r_w}\right)^{\frac{1}{2}} \sum_{i=1}^{coil} G(l_i) + \frac{\mu_0 \omega}{2\pi} \sum_{i=1}^{CV}\left(\frac{r_i}{r_w}\right)^{\frac{1}{2}} \sigma_i A_{I,i} S_i G(l_i)$$

$$A_I = -\frac{\mu_0 \omega}{2\pi} \sum_{i=1}^{CV}\left(\frac{r_i}{r_w}\right)^{\frac{1}{2}} \sigma_i A_{R,i} S_i G(l_i)$$

Here, $G(l_i)$ is a function of complete elliptical integrals and $CV$ indicates control volume [3]. The coil current $I$ appearing in the above expression is calculated from the specified total power dissipated into the discharge ($P_o$). The later can be defined as:

$$P_0 = \int_v Q_j(v')dv'$$

Here $Q_j$ is the local energy dissipation rate, as given in equation 26.

## 4. Numerical Procedure

A summary of torch geometry and operating conditions is given in figure-1 & table-1. Helium is mixed with Argon & injected through the nozzle $N_3$ as a sheath gas into the Argon plasma. Swirl is given either at the nozzle $N_2$ (= $w_2$ m/s) or at the nozzle $N_3$ (= $w_3$ m/s). In our calculations we have made the inlet swirling proportional to the inlet axial velocity ($u_1$) of the carrier gas at the nozzle $N_1$. We have calculated the control parameters with three values of swirling velocities ($w_2$ or $w_3$) amounting to $2u_1$, $4u_1$ and $6u_1$. Numerical calculations are performed both with and without Helium in the sheath gas. For calculations with Helium in the sheath gas, we have considered three inlet concentrations with Helium mass fraction amounting to $C_w = 0.1, 0.3$ and $0.5$.

The governing equations 1-4, 7-9, 13, 19 & 20, along with the boundary conditions are solved using the SIMPLER algorithm developed by Patankar [14]. A non-uniform grid of 55x35 points in the axial and radial directions and a staggered arrangement of variables are used for present calculations. The solution is considered converged when the variation of each field in successive iterations is less than 0.05%. For Benchmarking purpose, in figure 2a,b we have compared our velocity and temperature profiles with the ones obtained in ref [5] under similar conditions. Both the curves show qualitative resemblance. The slight difference in the profile can be attributed to the fact that the work in ref-[5] uses a 1-D electromagnetic formulation, while we are using 2D formulation, where the backflow is expected to be somewhat less.

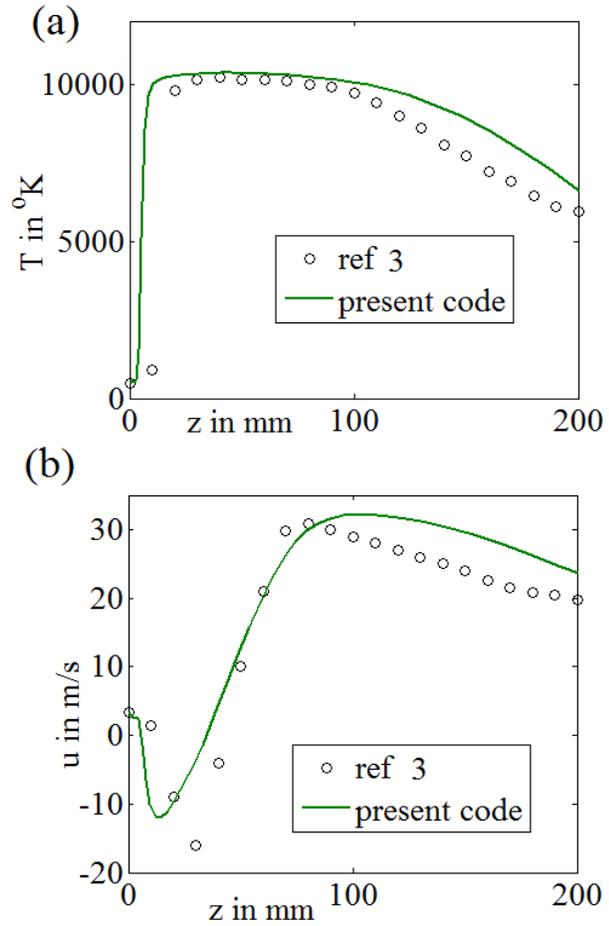

*Figure 2: Benchmarking curve comparing our results for (a) centerline temperature and (b) velocity with ref [5] for Q = 5L/min & w = 0. (The turbulence model is same as [5]. However, we have adapted a 2D model for the electromagnetic fields).*

## 5. Results and discussion:
### 5.1. Vector potentials, Joules heating and Lorentz forces

Figure 3a plots a gives the contour plots of $A_R$ (Real part of the vector potential) while figure 3b gives the contour plot of $A_I$ (Imaginary part of vector potential). As seen in the figure, the $A_R$ lies mostly near the tube wall, while $A_I$ moves somewhat inwards, towards the high-temperature regions. The confinement of $A_R$ near the tube wall results from skin effect that prevents the field from penetrating into the hot conducting plasma. $A_I$, is, on the other hand generated by the induced currents within the plasma, and its peak lies on the region where the induced current density is highest. As both the fields are out of phase and as $A_I$ increases with the increase of conductivity of the medium, $A_R$ gets confined to a decreasingly small region of the torch wall. In the typical case shown in figure 3, the maximum of $A_I$ occurs near $r = 15$ mm.

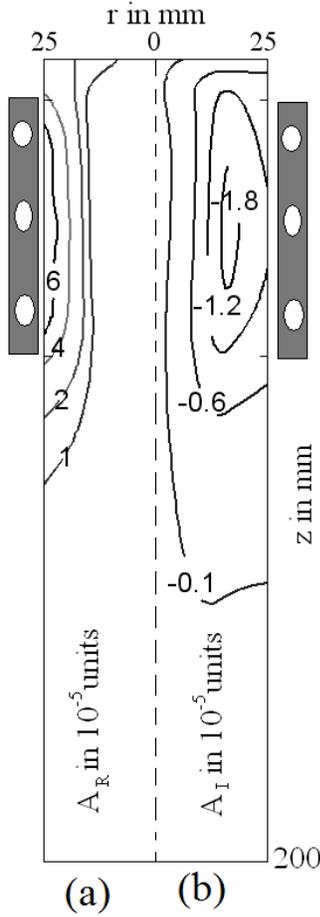

*Figure 3: Contour plot of AR and AI with swirling w2=2u and*

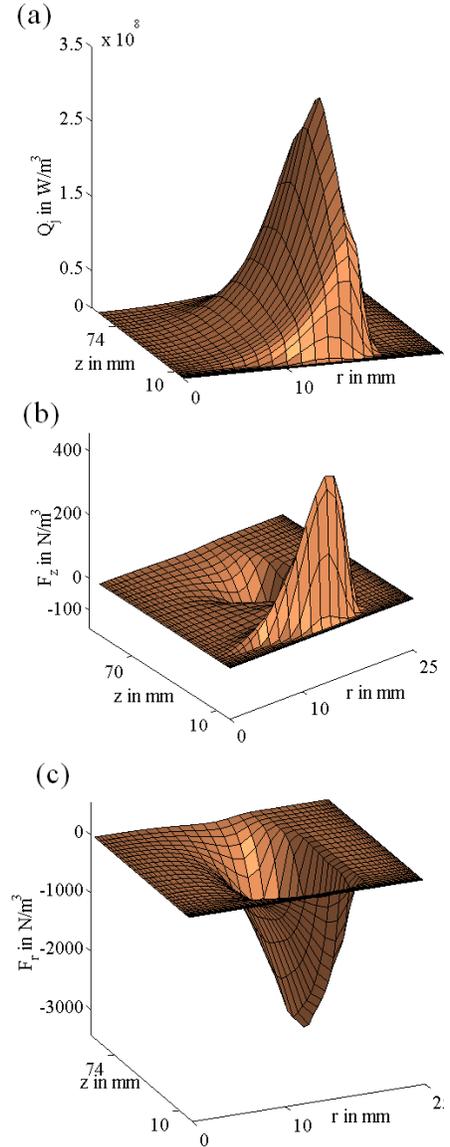

The effect of electromagnetic fields on the control characteristics of the plasma torch is seen in terms of the Joules heating and Lorentz forces affecting the flow & energy fields. Figure 4a-c gives the surface plot of $Q_j$ (Joules heating), $F_r$ (radial part of Lorentz force) & $F_z$ (axial part of Lorentz force) for the same operating conditions as described in figure-3. The joules heating $Q_j$ has its maximum near $r = 15$ mm, which is also the position where $A_I$ has its peak. The heating of plasma is thus almost entirely due to the induced fields generated through the inductive coupling with the external time-varying current. The radial Lorentz force $F_r$ is all-negative, and hence acts inward, towards the tube axis. $F_z$ is much smaller (generally by a factor of 5-

*Figure 4: Surface plot of Qj , Fz and Fr with swirling w2=2u and Cw=0.1.*

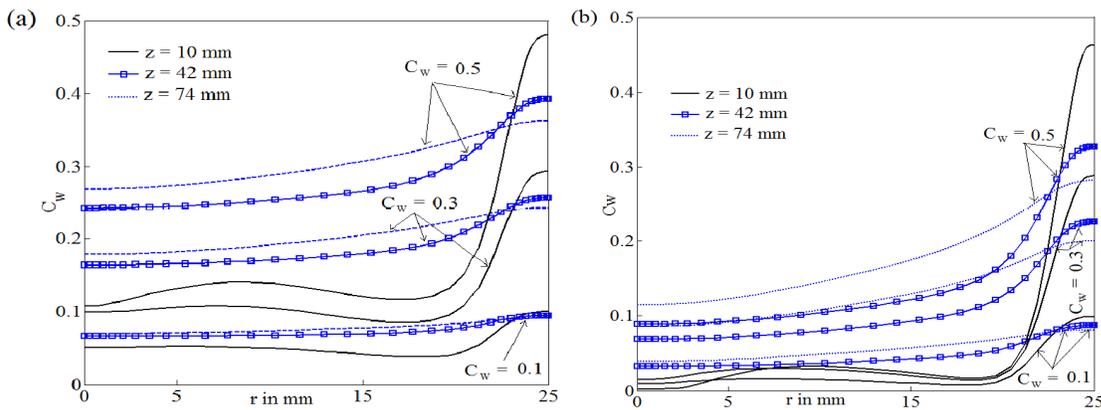

*Figure 5: Radial distribution of Helium concentration at the coil entrance (z =10 mm), coil centre (z = 42 mm) and coil exit (z = 74 mm) for (a) swirling at nozzle $N_2$ ($w_2 = 4u_1$) & (b) swirling at nozzle $N_3$ ($w_3 = 4u_1$).*

10) as compared to $F_r$, and has both negative as well as positive components. The effect of $F_z$ is in lowering the backflow caused by electromagnetic pumping on the upstream end of the coil, which is seen as the negative region in the centerline velocity curve (Figure 2b).

### 5.2. Influence of Swirling on diffusivity of Helium

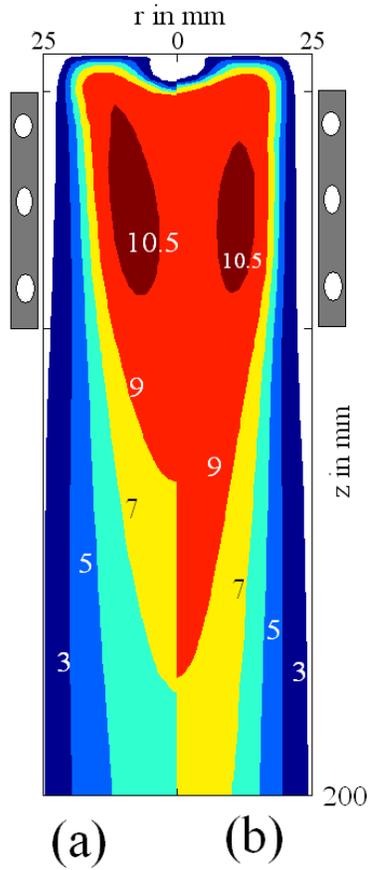

**Figure 6**: Contour plot of T in $x10^{3o}K$ with $w_2 = 2u_1$ and (a) $C_w = 0.5$ (b) $C_w = 0.1$.

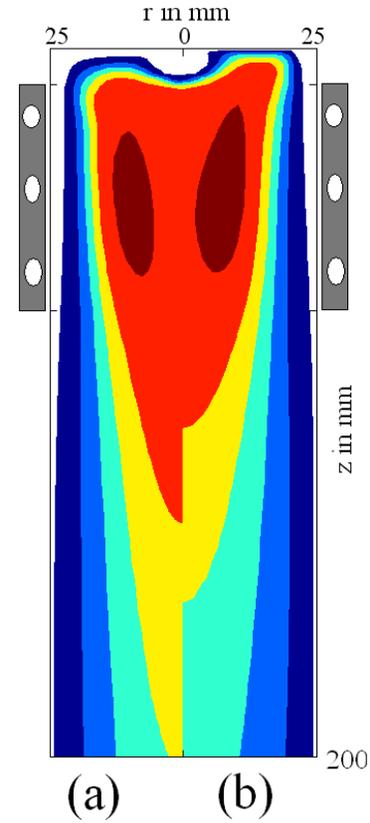

**Figure 7**: Contour plot of T in $x10^{3o}K$ with $C_w = 0.5$ and (a) $w_3 = 4u_1$ (b) ) $w_2 = 4u_1$

Figure 5a-b plots the distribution of Helium mass fraction at the beginning of the RF coil ($z = 10$ mm), at the middle of the coil ($z = 42$ mm) and at the end of the coil ($z = 74$ mm) for the same amount of swirling ( $= 4u_1$) given to the gas injected at Nozzle $N_2$ and $N_3$ respectively. Here the curves are plotted for three different Helium mass fraction at the inlet $N_3$, namely $C_w = 0.1, 0.3$ & $0.5$. As seen in the graphs, the concentration of helium at the middle and end of the coil region increases almost linearly with the increase in the injection rate of helium in both the cases. However, the concentration of helium at the central region is considerably higher when swirling is applied to the second nozzle ($N_2$). Helium diffuses considerably from the fringe to the central region by the swirling flow of $w_2$.

When swirling is given to gas in the 3$^{rd}$ Nozzle $N_3$(figure 5b), the concentration of Helium is found to be much less as compared to the earlier case shown in figure 5a. This is due to relatively low diffusion caused by swirling flow $w_3$. In other words, the mixing process of argon and helium and the diffusion of helium depend strongly on the position of the swirl.

### 5.3. Influence of Helium on Temperature profile.

Figure 6 plots the contour plot of temperature distribution with swirling $w_2 = 2u$ for two Helium concentrations, $C_w=0.1$ & $0.5$. Looking at the curves, we see that at the coil region, the high temperature (T > 10,500K) region is extended in case of higher injection of Helium (fig 6a). This is because with the diffusion of Helium into the plasma region, the electrical conductivity of plasma decreases. As we have kept the power dissipated to the plasma (i.e. sum of Joule heating over each CV) constant ($Po = 8kW$), coil current changes to keep the total power constant. Hence the general tendency observed in case of Joule heating is that with increase in the Helium concentration, although the maxima (peak) of the joules heating goes down, its base broadens and the profile diffuses to keep the total power $Po$ constant.

This increase in T in the central region is not carried to the torch exit and the gas quickly cools down to a much lower temperature due to intermixing of the two gases. In the particular case shown in the figure 6, the maximum exit temperature for figure 6a is found to be about 75% of maximum exit temperature of figure 6b.

Coming to the influence of swirling position on the temperature profile, from figure 7a-b, we see that temperature reduces considerably when swirling is given to Nozzle $N_2$, as compared to when it is given to the Nozzle $N_3$. This is because with swirling given to $N_2$, Helium diffuses in a much greater way to the central regions cooling the plasma gas there.

### 5.4. Turbulence Characteristics
The flow in RF plasma torch is characterized by low velocity and high temperature, and the flow is considered laminar when laminar viscosity is nearly of the same order as turbulent viscosity.

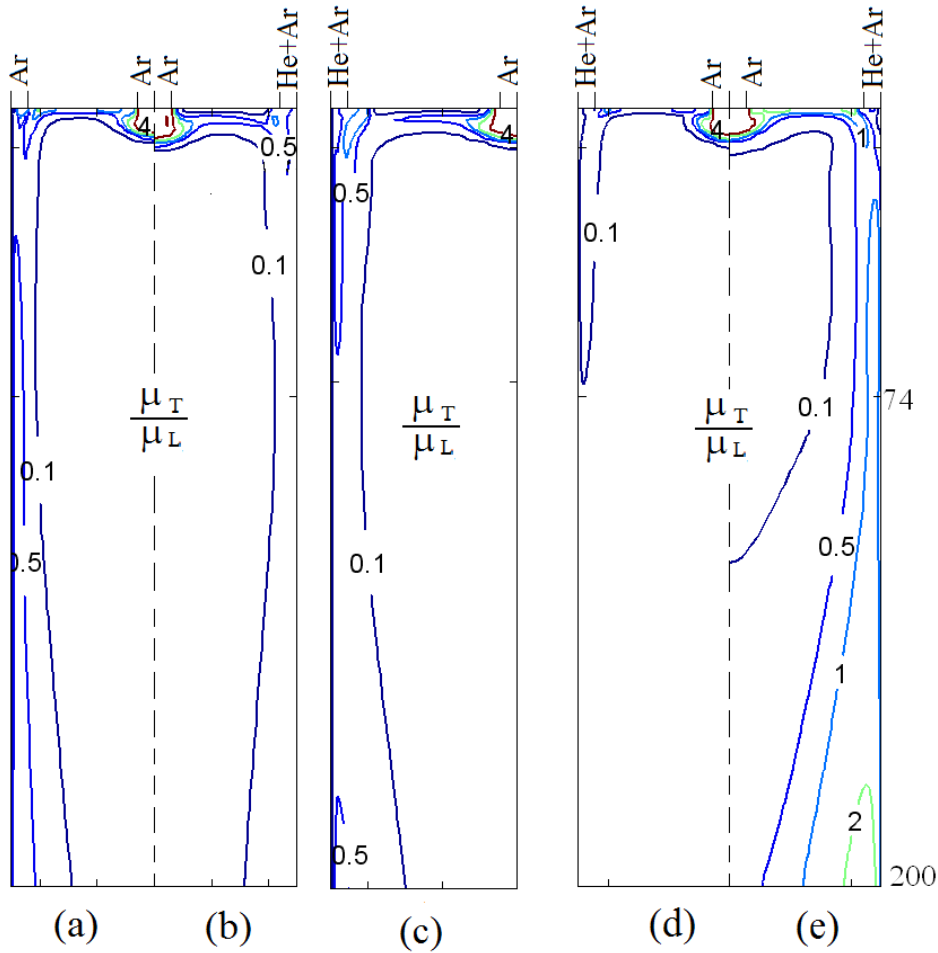

Figures 8a-e plots the contours for the ratio of turbulent to laminar viscosities $\left(\frac{\mu_T}{\mu_L}\right)$ for different flow conditions. It is found that the turbulent regions in our case are mainly confined to the gas inlet and near the wall of the tube. Other regions are found to be predominantly laminar. The turbulence is largest for the case without injection of Helium (figure 8a), while with the increase in injection rate, the amount of turbulent regions goes down. The near wall turbulence increases with increase in swirl (figures 8b&c). Also, turbulent viscosity is relatively higher when swirling is given to the nozzle $N_3$ then when it is given to nozzle $N_2$.

**Figure 8:** Ratio of Turbulent viscosity to laminar viscosity for (a) $C_w = 0$, $w_3 = 2u_1$ (b) $C_w = 0.5$, $w_3 =2u_1$ (c) $C_w = 0.5$, $w_3 = 6u_1$ (d) $C_w = 0.5$, $w_2 =4u_1$ and (e) same as (d), but with standard k-$\varepsilon$ model [15] applied instead of the RNG k-$\varepsilon$ model.

Finally, the contours in figure 8d &e plots the laminar to turbulent viscosity ratio as obtained with both RNG $k$-$\varepsilon$ model as well as with the standard $k$-$\varepsilon$ model for the same external conditions. It is found that the turbulence as calculated with the standard standard $k$-$\varepsilon$ model is much higher as compared to the RNG $k$-$\varepsilon$ model, especially in the hotter regions.

### 6. Conclusion:
In this paper we have studied the influence of injection of secondary gas (Helium) on the control characteristics of plasma torch. We have studied how the swirling given at different inlet affects the plasma parameters within the torch. Finally we have studied the turbulence within the torch leaning on the RNG $k$-$\varepsilon$ model of turbulence. It is generally observed that when helium is injected into the plasma, the high temperature region reduces. The amount of this reduction depends on how much helium diffuses to the central region. The diffusion is largest, when swirling is given to sheath inlet $N_3$.

Turbulence effects decrease with the increase in injection rate of Helium. Also, the turbulence calculated with the present model (RNG $k$-$\varepsilon$ model) is found to be less compared to the standard $k$-$\varepsilon$ model.